\documentclass[12pt]{article}
    
\begin{document}
        
\hoffset=-.5in

\voffset=-1in

\def\a-U{$\alpha$--U}

\def\Tc{$T_c$ }

\def\TN{$T_N$ }    

\def\Hc2(T){$H_{c2}$(T) }

\vspace*{.8in} 

\centerline{\large\bf Magnetotransport and Superconductivity of $\alpha$--Uranium} 

\vspace*{.6in}
    
\centerline{G.~M. Schmiedeshoff, D. Dulguerova, J. Quan, and S. Touton} 
    
\smallskip\centerline{\it Department of Physics, Occidental College, Los Angeles, California 90041, USA} 

\bigskip 

\centerline{C.~H. Mielke, A.~D. Christianson and A.~H. Lacerda} 
    
\smallskip\centerline{\it National High Magnetic Field Laboratory, Pulse Facility,}

\smallskip\centerline{\it Los Alamos National Laboratory, Los Alamos, New Mexico 87545, USA} 

\bigskip 

\smallskip\centerline{E. Palm, S.~T. Hannahs, and T. Murphy}

\smallskip\centerline{\it National High Magnetic Field Laboratory, Tallahassee, Florida, 32310, USA}

\bigskip

\centerline{E.~C. Gay and C.~C. McPheeters} 
    
\smallskip\centerline{\it CMT Division, Argonne National Laboratory,}

\smallskip\centerline{\it Argonne, Illinois 60439, USA} 

\bigskip 

\centerline{D.~J. Thoma, W.~L. Hults, J.~C. Cooley, A.~M. Kelly, R.~J. Hanrahan, Jr. and J.~L. Smith} 
    
\smallskip\centerline{\it Los Alamos National Laboratory, Los Alamos, New Mexico 87545, USA}

\bigskip 

%\centerline{(Version: \today)} 
\centerline{(Received: 2 May 2003)} 

\medskip\centerline{(Accepted: 4 January 2004)}

\baselineskip=16pt

\bigskip

\noindent{\bf ABSTRACT} 
    
\bigskip\noindent We have measured the electrical resistivity, magnetoresistance, and Hall effect on several new single crystal samples and one polycrystalline sample of \a-U.  The residual resistivity ratios of these samples vary from 13 to 315.  Matthiessen's law appears to hold above the onset of the charge density wave phase transitions that begin near 43 K, but not  below this temperature.  Sharp features at all three charge density wave transitions are observed and the effects of high magnetic fields on them are presented and discussed.  The magnetoresistance is anisotropic, reaches 1000\% at 2 K and 18 T, and does not exhibit Kohler scaling.  The Hall coefficient is positive, independent of magnetic field, and slightly temperature dependent above about 40 K in agreement with earlier studies.  Below 40 K the Hall coefficient changes sign as the temperature falls, varies with field, and becomes much more strongly negative at the lowest temperatures than has been reported.  Some of our results suggest that a spin density wave may coexist with the charge density wave states.  Superconductivity is observed in two of our samples, we argue that it is intrinsic to \a-U and suggest that it is consistent with a two--band model.  Several parameters characterizing the transport and superconductivity of \a-U are estimated.  

\bigskip
    
\section{Introduction}

``One would think that the condensed-matter physics of the elements would have been understood many years ago, but there are still a number of elements such as manganese, chromium, and, of course, uranium, that have complex structural or magnetic behavior; much of which is still the subject of research.  The studies of these materials have usually spanned several decades.  Often progress in understanding comes to a standstill until a new idea emerges, often unrelated to the problem in question, but soon seen to apply to it, and then progress occurs'' \cite{Lander} (hereafter LFB).  By 1994, progress in understanding the physics of the alpha phase of metallic uranium had come to a virtual halt due to the inability to fabricate high quality single crystals.  The incentives to create single crystals and the difficulties in doing so are covered in the well written and comprehensive review article of LFB quoted above.

\medskip Just a few years later, a significant advance in single crystal fabrication was made at Argonne National Laboratory by a group developing a new technique to electrometallurgically extract pure uranium from spent nuclear fuel \cite{McPheeters}:  the extracted uranium emerged from the melt as a crystalline cluster (a photograph of which appears on the cover of the journal containing \cite{McPheeters}).  The samples we report on in this paper were cut from this cluster.

\medskip Solidifying into the $\gamma$ phase at just over 1100 $^o$C at ambient pressure, uranium undergoes structural phase transitions into the $\beta$ phase just under 800 $^o$C and then into the $\alpha$ phase near 660 $^o$C.  The structure of \a-U is well characterized as ``complex'' in the literature \cite{Kittel}, having an orthorhombic structure reminiscent of corrugated cardboard when viewed along the [100] axis.  As one would expect, most of the physical properties of \a-U are anisotropic.  In the electrical resistivity as measured along the [100] and [010] axes for example, an anisotropy of about 1.6 decreases to about 1.2 as the temperature falls from room temperature to 4.2 K \cite{Berlincourt_1959}. 

\medskip Surprisingly for a three--dimensional metal and a non--magnetic element, \a-U passes through a series of three charge density wave (CDW) phase transitions as the temperature falls.  The {\bf q} vector of the charge density wave may be written as

\begin{equation} 
{\bf q}_{CDW} = q_x{\bf a^*} + q_y{\bf b^*} + q_z{\bf c^*},
\end{equation}

\medskip\noindent where {$\bf a^*$}, {$\bf b^*$}, and {$\bf c^*$} are the reciprocal lattice vectors, with magnitudes $2\pi/a$, $2\pi/b$, and $2\pi/c$ respectively, of the real space lattice vectors {\bf a}, {\bf b}, and {\bf c}.  The warmest of the CDW phases, $\alpha_1$, is an incommensurate CDW state which appears below about 43 K and is characterized by $q_x=\pm{0.490}$, $q_y=\pm{0.131}$, and $q_z=\pm{0.225}$.  In the $\alpha_1$ phase $q_z$ decreases linearly with decreasing temperature while $q_y$ and $q_x$ increase.  In the $\alpha_2$ phase below 37 K, $q_x$ locks into a value of 1/2, and in the $\alpha_3$ phase appearing below 22 K, $q_y$ and $q_z$ lock in to values of 1/6 and 5/27 respectively. The CDW is thus fully commensurate with the lattice in the $\alpha_3$ phase \cite{Marmeggi_1990}.  First-principles calculations show that the $\alpha_1$ state is due to a Peierls-like transition opening partial gaps at the Fermi level, these regions of the Fermi surface show a strong nesting of fairly narrow $f$-bands \cite{Fast_1998}.

\medskip  It would not be difficult to argue that the superconductivity exhibited by \a-U is the most interesting and unusual of any of the elements if one could be sure that it was intrinsic.  The history of the superconductivity observed in \a-U samples is rich and well documented by LFB:  Over the last 70 years superconducting critical temperatures \Tc have been reported that vary from below 0.1 K to about 1.3 K with the more impure samples having the higher \Tc\negthinspace.  Unlike most of the elements, magnetic measurements show that \a-U is a Type II rather than a Type I superconductor.  The large enhancement of \Tc by hydrostatic pressure (0.18 K/bar) was the largest known at the time it was first measured.  

\medskip Bulk superconductivity with a BCS character was confirmed by specific heat measurements at high pressure, but not at ambient pressure (especially for single crystal samples) where measurements of the Meissner effect led to estimates of a superconducting fraction of about 10\%.  As a result, there has been a long--standing debate over whether ambient pressure superconductivity is intrinsic to \a-U or due to some kind of filamentary and/or impurity phase, the $\beta$ and/or $\gamma$ phases of uranium being the primary suspects.  The argument, briefly stated, is that these higher temperature phases of uranium are formed as the samples are cooled from the liquid phase and are stabilized (at grain boundaries for example) by the stresses caused by the large anisotropic thermal expansion.  Polycrystals would therefore be more likely to show the characteristics of bulk superconductivity than single crystals, and the superconducting transitions, reflecting the range of stresses at the grain boundaries, would be rather broad as reported by many researchers.  There has been little work reported on the superconductivity of \a-U since the existence of the CDW phases has been confirmed. Since both BCS superconductivity and CDW's gap the Fermi surface one might expect them to ``compete'' for the available area.  While many of the unusual features of the superconductivity can be understood given the existence of CDW's, it has not been conclusively demonstrated that the superconductivity observed in \a-U at ambient pressure is intrinsic.

\medskip In this paper we report on measurements of the electrical resistivity, magnetoresistivity, and Hall effect on new high quality samples of \a-U.  A small portion of the results we report below has appeared elsewhere \cite{JLS1,Lashley_2001}.  So much earlier work has contributed to our present understanding of \a-U that it is not practical for us to attempt a comprehensive comparison with earlier studies.  Fortunately, as mentioned above, the review article by LFB is an excellent guide to the literature, and we encourage the reader to consult it.

\section{Experimental Details}

Uranium single crystals were grown by electrotransport through a molten salt bath of LiCl--KCl eutectic containing about 3\%  UCl$_3$ by weight.  The uranium formed dendrites on a stainless steel cathode in the shape of parallelogram--edged platelets.  Since the deposition temperature was lower than that of the $\alpha$--$\beta$ structural transition, these platelets are nearly perfect strain--free single crystals.  A back--reflection Laue technique showed that the crystalline [001] axis was normal to the plates, however the orientation of the other crystalline axes is presently uncertain for our crystals.  Our transverse magnetoresistance data are therefore presented with the magnetic field either parallel or perpendicular to the [001] axis.  A polycrystal sample was made by arc melting crystals grown in the manner described above.

\medskip The electrical resistance was measured using a standard four--probe ac technique at 16 Hz.  Platinum leads were affixed to the single crystals with silver epoxy and to the polycrystal sample by spot welding.  The current was always applied perpendicular to the c--axis.  Where possible, the resistivity was calculated from the dimensions of the sample with an estimated absolute uncertainty of about 20\%.  The temperature was determined from Cernox thermometers using the manufacturers calibrations for measurements in $^4$He or $^3$He cryostats, or by a RuO$_2$ thermometer calibrated against the melting curve of $^3$He for measurements in a dilution refrigerator.

\medskip The Hall resistance was measured on single crystal Sample 3 (see below) which was ground to a thickness of $55{\pm}10$ $\mu$m.  The resistance was measured with voltage leads aligned perpendicular to the current, the Hall resistance $R_{Hall}(T, H)$ was then determined from:

\begin{equation}
R_{Hall}(T, H) =  \frac{1}{2} [R(T, H) - R(T, -H)],
\end{equation}

\medskip\noindent to remove any spurious magnetoresistive contribution.

\medskip In this paper we describe measurements on four single crystals and one polycrystal of \a-U, some characteristics of these samples are gathered in Table 1.   We determined the residual (or impurity limited) resistivity ratio 
($RRR = R$($\sim$300 K)/$R$($\sim$2 K) where $R(T)$ is the electrical resistance at temperature $T$).  $RRR$'s for our single crystal samples range from  66 to 315.  We have been unable to find $RRR$ data in the literature larger than about 36 (LFB) which suggests that our crystals are of very high quality. The $RRR$ for our polycrystal (13) is comparable to that of earlier measurements in the literature.

\section{Results}

\subsection{Resistivity}

The reduced resistivities ($\rho(T)/\rho$(300 K)) of three of our samples are shown in Fig. 1a. Fig. 1b shows the low temperature region of the data with the curves for samples 1 and 5 shifted vertically by the amounts shown in parentheses.  The resistivity exhibits negative curvature from room temperature to the onset of the CDW states below 50 K and the more common positive curvature below the transitions.  When possible we fit the resistivity between 60 K and 200 K and between 4 K and 10 K to the form

\begin{equation}
\rho(T)=\rho_o+AT^n, \label{eq:rho_of_T}
\end{equation}

\medskip\noindent the resultant values of $n$ are listed in Table 1.  The mean value of $n$ (excluding the ``bent'' sample, see below) is 0.70$\pm$0.06 for the higher temperature region and 3.2$\pm$0.2 for the low temperature region.  The gradual drop in resistivity associated with the onset of the CDW states is barely noticeable in the polycrystal sample.  

\medskip The low temperature electrical resistivity of sample 1, in both zero and high magnetic fields applied along the $c$-axis, is shown in Fig. 2. Sharp features at temperatures $T_1$, $T_2$, and $T_3$ mark the transitions into the charge density wave states  $\alpha_1$, $\alpha_2$, and $\alpha_3$ respectively.  The application of magnetic fields to 18 T introduces a large magnetoresistive component but does not affect the temperatures of the CDW features within experimental resolution.  To locate these features we use the temperature derivative of the resistance of sample 2 (for example) as shown in Fig. 3. (Measurements with an 18 T field applied perpendicular to the c--axis are identical, within experimental resolution, to the 18 T data shown.)  If we identify the onset of the $\alpha_1$ phase with the sharp upturn in $dR/dT$ then $T_1 =$ 43.5 K, if we identify the onset of the $\alpha_2$ phase with the deep minimum in $dR/dT$ then $T_2 =$ 37.5 K, and if we identify the onset of the $\alpha_3$ phase with the maximum in $dR/dT$ in the warming branch of the data then $T_3 =$ 23.4 K.  These temperatures are all in good agreement with earlier evaluations cited in LFB and with the more recent results of \cite{Lashley_2001}.

\medskip If we attribute the general downward shift of the 18 T data in Fig. 3 to magnetoresistivity effects, then the shape of the CDW transition at $T_1$ does not appear to be significantly affected by high magnetic fields, but the transitions at $T_2$ and $T_3$ are:  The minimum at $T_2$ deepens as the field increases and the maximum in the cooling branch of the data near 20.5 K is much more pronounced at 18 T than in zero field.  These features are significantly less prounounced in polycrystal sample 5 as shown in Fig. 4.  The small hysteresis in the resistivity (Fig. 4a) is not seen above the noise in the first derivative (Fig. 4b).  Most of the structure associated with the CDW transitions in the single crystal samples is absent in the polycrystal, but the maximum in d$\rho$/d$T$ at 39.5 K correlates with a similar feature from single crystal Sample 2 (at about 40 K, see Fig. 3).  

\medskip The thermal hysteresis in the electrical resistivity seems most closely associated with the $\alpha_3$ phase.  The size of the thermal hysteresis of single crystal Sample 1 is shown in Fig. 5 where we define

\begin{equation}
\frac{\Delta\rho}{\rho_{ave}} = 2\enspace\frac{\rho_-(T) - \rho_+(T)}{\rho_-(T) + \rho_+(T)}. \label{eq:hysteresis}
\end{equation}

\medskip\noindent Here $\rho_-(T)$ and $\rho_+(T)$ represent the resistivity when cooling or warming the sample respectively.  High magnetic fields reduce $\Delta\rho/\rho_{ave}$ while the maximum broadens somewhat and moves to higher temperatures.  There is a shoulder in the zero field data near 22 K.  This shoulder corresponds more closely to the maximum hysteresis of two other samples as shown in Fig. 6 where we have plotted $\Delta\rho/\rho_{ave}$ in zero-field for single crystal Sample 2 and polycrystal sample 5 as well as that of single crystal Sample 1.  Note that the hysteresis changes sign at the lowest temperatures for our highest quality crystal and that the maximum in $\Delta\rho/\rho_{ave}$ is very broad for the polycrystal.

\medskip The reduced resistivity of single crystal Sample 3 and its derivative with respect to temperature are shown in Figs. 7a and 7b respectively.  This is a very thin sample ($55{\pm}10$ $\mu$m) used to study the Hall effect discussed below.  In the midst of measuring this sample a lead fell off and had to be replaced, a process which caused a moderate amount of sample bending.  While the Hall resistance of this sample was unaffected by the bending within experimental uncertainty, the residual resistivity ratio dropped from 183 to 29 and the curvature in the normal state (above 60 K) became less pronounced (see Fig. 7a and Table. 1).  No thermal hysteresis was visible in this sample either before or after bending, and the features at the CDW transitions are quite indistinct in comparison to our other single crystal samples.  The maximum in d$R$/d$T$ shown in Fig. 7b is smaller in this bent sample, but its temperature (40 K) is unaffected and very close in temperature to a similar feature in our other \a-U samples.

\subsection{Magnetoresistivity}

The transverse magnetoresistivity, given by

\begin{equation}
\frac{\Delta\rho}{\rho} = \frac{\rho(H) - \rho(0)}{\rho(0)},
\end{equation}

\medskip\noindent is shown in Fig. 8 for single crystal Sample 2.  At temperatures above about 40 K the magnetoresistance is roughly quadratic with field as expected for a simple metal.  Near 20 K, in the midst of the CDW phase transitions, the magnetoresistance exhibits a pronounced negative curvature. At lower temperatures near 2 K the magnetoresistance is again roughly quadratic (compare, for example, the 2 K and 40 K data in Fig. 8b).  We see no sign of the magnetoresistance tending toward saturation in high fields nor any sign of oscillatory behavior from the Shubnikov -- de Haas effect that one might expect for high quality crystals.

\medskip If there is a single species of charge carrier and if the scattering time $\tau$ is the same at all points on the Fermi surface (for any Fermi surface geometry) then the magnetoresistance should be proportional to $\omega_c\tau$ where $\omega_c$ is the cyclotron frequency \cite{Pippard}.  This behavior, known as Kohler's law, can be demonstrated by plotting $\Delta\rho/\rho$ vs. $H/\rho(0)$.  The failure of Kohler scaling for \a-U is graphically demonstrated in Fig. 9 where the intermediate temperature data does not fall on the curve delineated by the high or low temperature data.

\medskip Similar data for polycrystal Sample 5 are shown in Fig. 10 with the field dependence of the magnetoresistivity shown in Fig. 10a and a Kohler plot of the same data in Fig. 10b.  The polycrystal magnetoresistivity is about 20 times smaller than that of the single crystal (similar to the relative size of their $RRR$'s).  The negative curvature exhibited by the single crystal's magnetoresistivity at intermediate temperatures is absent in the polycrystal.  And, while Kohler scaling fails for both samples, it does not fail the same way: the low temperature Kohler scaled data of the polycrystal does not fall near that of the high temperature data as it does for the single crystal.

\subsection{Hall Effect}

Isothermal measurements of the field dependent Hall resistance ($R_{HALL}$) in fields to 9 T applied parallel to the c--axis are shown in Fig. 11a.  We find $R_{HALL}$ to be positive and linear in $H$ from 100 K down to about 35 K but it becomes quite nonlinear with field below this temperature in addition to becoming strongly negative at the lowest temperatures studied.  A portion of the low field data is shown in Fig. 11b, the change in the sign of the slope, $dR_{HALL}/dT$, occurs near 22 K in zero field, close to $T_3$.   

\medskip The Hall coefficient is given by $r_H \equiv -1/ne$, where $n$ is the number density of the charge carriers and $e$ is the magnitude of the electronic charge.  In the usual way this reduces to

\begin{equation}
r_H = R_{HALL}\frac{t}{H},
\end{equation}

\medskip\noindent where $t = 55\pm10$ $\mu$m is the thickness of single crystal sample 3.  Hall coefficients derived from our measurements are shown in Fig. 12 (where the uncertainty bars do not incorporate the 18\% uncertainty in $t$ for clarity).  At temperatures above about 35 K where $R_{HALL}(H)$ is linear, $r_H$ is determined from the slope of the data and plotted as solid squares in Fig. 12.  The other points in Fig. 12 were determined by holding $T$ and $H$ fixed, averaging $R_{HALL}$ for an appropriate period of time, reversing $H$, repeating the operation, and then applying Eqn. 2.  The solid lines in the body of Fig. 12 are guides to the eye.  

\medskip  We find that $r_H$ decreases slightly with decreasing temperature from 100 K, passes through a broad, shallow minimum near 50 K and then over a somewhat sharper maximum near 35 K.  Below 35 K $r_H$ drops precipitously and acquires a field dependence. The inset shows the field dependence of the Hall coefficient at 5 K, the solid line is a quadratic fit to the data which extrapolates to (-$58.7{\pm}6$) $\times{10^{-11}}$ m$^3$/C and exhibits a broad minimum centered near 6.4 T.

\subsection{Phase Transitions Below 5 K}

Evidence for phase transitions below 5 K in single crystal sample 4 and polycrystal sample 5 is presented in Fig. 13 which shows the temperature dependence of the resistance of these samples in zero field.  The drop in resistance, below 1 K and 2 K for samples 4 and 5 respectively, is identified as the onset of superconductivity.  Broad resistive superconducting transitions such as these are not uncommon in \a-U \cite{Lander} and may reflect sample inhomogeneity, however, the shape of these transitions is also consistent with flux flow effects, effects which can be enhanced by coupling to the CDW \cite{Chung_1994}.  For further analysis, we identify the superconducting transition temperature $T_c$ as the temperature at which the resistance has dropped to 90\% of its extrapolated normal state value.  The inset of Fig. 13 shows the isothermal resistance of polycrystal Sample 5 as a function of field.  The upper critical field at $H_{c2}$, determined using a similar 90\% criteria, and a second feature at $H_o$ are identified with arrows.

\medskip The upper critical fields \Hc2(T) of Samples 4 and 5 are plotted in reduced form ($T/T_c$ vs. $H_{c2}(T)/H_{c2}(0)$) in Fig. 14 where the solid line is a guide to the eye.  Parameters describing the superconducting properties of these two samples are gathered in Table 2. $T_c$ is determined from a linear fit of the low field data by evaluating the fit at $H=0$, the slope of this fit gives $(dH_{c2}/dT)|_{T_c}$.  $H_{c2}(0)$ is determined by plotting the low temperature data vs. $T^2$, fitting this data to a straight line and evaluating the fit at $T=0$.  

\medskip Also shown in Fig. 13 is a second feature in the resistance of polycrystal sample 5 near 3 K.  This feature is identified as $T_0$ in the body of the figure and as $H_0$ in the inset. $H_0$ increases as the temperature falls as shown in Fig. 15 where the solid line is a power--law fit of the data to the form

\begin{equation}
T_0(H) = T_0(0) - {\beta}H^n.
\end{equation}

\medskip\noindent We find $T_0(0)$ = 3.07 $\pm$ .03 K, $\beta$ = 159 $\pm$ 15 mK/(T)$^n$, and $n$ = 1.34 $\pm$ .04.  The inset of Fig. 15 shows the temperature dependence of the resistivity of polycrystal Sample 5 in a field of 0.5 T, a field strong enough to suppress superconductivity in this sample.

\section{Discussion}

\subsection{Resistivity}

\noindent If the scattering from impurities is independent of the scattering from phonons and if the relaxation time is isotropic, the resistivity should be separable into two terms: a constant (or residual) resistivity due to the impurities and a temperature dependent term due to the phonons.  This behavior is known as Matthiessen's Law.  The clustering of the resistivity curves above 60 K, shown in Fig. 1a, suggests that Matthiessen's Law holds (at least qualitatively) in the normal state of \a-U, but the divergence of the curves at lower temperatures suggests that it does not hold in the region of the CDW states.  Since the CDW transitions modify an already anisotropic Fermi surface, this result is not unexpected.  At the lowest temperatures we find that a $T^3$ temperature dependence is just at the edge of our experimental uncertainty for all samples (see Table 1).  However, if the variance in the low temperature exponent is intrinsic (as suggested by \cite{Arjas_1964}), then Matthiessen's Law fails at the lowest temperatures too.  

\medskip The data of Fig. 1b show that the general trend of the temperature dependence is for $\rho(T)$ to drop as the CDW state is approached; this tendency is more pronounced in the better samples. This is somewhat surprising since a CDW is characterized by a gap opening up at the Fermi surface and the presence of such a gap should lead to higher resistivities.  For example, the CDW compound NbSe$_3$ shows a resistivity that generally decreases with falling temperature with upward jumps at its CDW transitions \cite{Gruner_1988}.  We suspect that magnetic fluctuations and/or anisotropy may play a role in this behavior. 

\medskip Negative curvature in the temperature dependence of the resistivity of \a-U has most recently been reported by \cite{Brodsky_1969} and by \cite{Sahu_1993}.  Whereas our higher temperature data appear to be well fit by a single power-law with an average value for the exponent near 0.7 (see Table 1), \cite{Brodsky_1969} report a power that varies from 1.3 at 42 K to 0.8 at 300 K.  \cite{Sahu_1993} have shown that the negative curvature continues up to the $\alpha-\beta$ structural phase transition and that higher pressures accentuates the curvature.  Various mechanisms for this unusual temperature dependence have been considered by \cite{Sahu_1993} who suggest that its origin lies in antiferromagnetic spin fluctuations of the 5$f$-electrons.  One might then expect a transition to antiferromagnetic order at low temperatures or an influence of such fluctuations on the thermodynamics.  Antiferromagnetism, however, has not been observed, and the low temperature specific heat does not contain the $T^3\log{T}$ term one would expect from the presence of spin fluctuations \cite{Lashley_2001}.  On the other hand, it may be that such fluctuations exist but at an energy too high to affect the low temperature behavior.  Such a situation may exist in $\alpha$-Ce \cite{Manley_2002}.

\medskip Another possible origin for the negative curvature observed above 60 K might be one of the interactions leading to non--Fermi liquid (nFL) behavior. Several systems in this increasingly broad category of $d$- and $f$-electron metals exhibit negative curvature in $\rho$(T), though, for analysis purposes, one would prefer it not be interrupted by phase transitions into either a CDW or superconducting state.  Such nFL behavior can originate if a system is near a quantum critical point (a $T=0$ antiferromagnetic phase transition for example), if it exhibits multi-channel Kondo interactions, or through a type of disorder characterized by a spread of Kondo temperatures \cite{Cox_1998,Stewart_2001}.  One might expect a quantum critical point in \a-U since the application of pressures of order 1 GPa suppresses the CDW states to $T=0$. A characteristic temperature for nFL behavior can be derived from the temperature dependence of the resistivity \cite{Stewart_2001} which is given by 

\begin{equation}
T_{nFL} = (A/\rho_o)^{1/n}, \label{eq:nFL}
\end{equation}

\medskip\noindent where the parameters are those of Eqn. \ref{eq:rho_of_T}.  This T$_{nFL}$ (which would be roughly equivalent to the Kondo temperature in such systems) ``gives an idea of the energy scale for the electron--electron interactions causing the non-Fermi liquid behavior'' \cite{Stewart_2001}.  Fitting Eqn. \ref{eq:nFL} to our high temperature data results in unphysical, negative values of $\rho_o$.  One might reasonably argue that this fact invalidates an nFL analysis.  Nevertheless we proceed by fitting the resistivity data below about 7 K to the form of Eqn. \ref{eq:rho_of_T} with $n=3$ and using the resultant value of $\rho_o$ in Eqn. \ref{eq:nFL} along with $A$ and $n$ deduced from the high temperature fits.  The results, which are not physically unreasonable, are shown in Table 1.  Unlike the values of $n$, T$_{nFL}$ correlates rather well with sample purity (assumed to scale with $RRR$) suggesting that the ``disordered Kondo'' flavor of non-Fermi liquid theory might also be a profitable approach in understanding the normal state of \a-U.  We also observe that these values of T$_{nFL}$ crudely span the temperature region from the onset of the CDW states to the onset of superconductivity.  It is difficult to imagine, however, that such low characteristic temperatures could significantly influence the behavior of $\rho(T)$ near the $\alpha-\beta$ phase transition where negative curvature in $\rho(T)$ continues to exist \cite{Sahu_1993}.  On the other hand, the resistivity measurements reported by \cite{Sahu_1993} show that the negative curvature is enhanced under pressure.  One might expect such behavior if high pressures push \a-U closer to a quantum critical point as suggested above.

\medskip Lastly we note the change in the resistivity due to the bending of single crystal sample 3 shown in Fig. 7.  Presumably this is driven by electronic scattering from dislocations generated in the bending process.  The data contained in Table 1 shows that the temperature dependence of the resistivity is also affected.  What kinds of dislocations are responsible? Indeed, we were quite surprised to find that these new crystals of \a-U bend easily and repeatedly without work hardening or necking when it has been our experience that actindes are generally rather tough.  We believe that
twinning and the running of twins over millimeters may be responsible, and we are presently testing this hypothesis. 

\subsection{Features associated with the CDW phase transitions}

\medskip Features at $T_1$ and $T_2$ appear in the resistivity measurements of \cite{Brodsky_1969} who also report thermal hysteresis with a maximum value of about 4\% near $T_3$.  By comparison, the features in our single crystal samples 1 and 2 are significantly more distinct (see Figs. 2 and 3), the thermal hysteresis is larger (10-15\%) and has a clear onset at $T_3$ (the hysteresis reported by \cite{Brodsky_1969} is similar to the relatively small, broad hysteresis we observe in our polycrystal sample 5, see Fig. 6).  

\medskip The sharpness of the CDW features does not directly correlate with the quality of our samples as characterized by their $RRR$ values.  This can be seen in Fig. 1b where the features of sample 1 are more distinct than those of either sample 2 (higher $RRR$) or sample 5 (lower RRR).  This may be due to the CDW state being more sensitive to impurities than the normal state as it is in the CDW compound NbSe$_3$ for example \cite{Ong_1979}.  

\medskip Magnetic fields do not affect the individual features consistently as can be seen in Figs. 2 and 3.  The feature at $T_1$ is relatively unaffected by high magnetic fields whereas the feature at $T_2$ evolves from a plateau-like feature to a distinct minimum, qualitatively similar to the much larger features at the CDW transitions in NbSe$_3$ \cite{Gruner_1988}.  High magnetic fields depress the magnitude of the thermal hysteresis that sets in near $T_3$ as shown in Fig. 5.  The shoulder near 22 K in zero field suggests that there may be two overlapping maxima in this sample, the maximum associated with the shoulder being less sensitive to the external field.  The thermal hysteresis of three samples are compared in Fig. 6 where it appears that the maximum of sample 2 coincides with the shoulder of sample 1.  Note also, in Fig. 5, that the thermal hysteresis appears to be changing sign as the temperature falls in high fields.  This tendency is clearly present in the thermal hysteresis of sample 2, shown in Fig. 6, where the sign change occurs near 10 K.  

\medskip Thermal hysteresis is common in CDW systems where it is associated with the pinning of the CDW by impurities.  For the thermal hysteresis of the resistivity in these systems, the warming trace always exhibits a higher value than the cooling trace (see the work on blue bronze by  \cite{Xue_1996} for example), just the opposite of what we generally observe for \a-U (except for the high field, low temperature limit of sample 1 shown in Fig. 5, or below 10 K for sample 2 as shown in Fig. 6).  We do not have an explanation for this unusual behavior.  

\medskip As with the sharpness of the featurs at the CDW phase transitions, $\Delta\rho/\rho_{ave}$ does not appear to correlate with $RRR$ as shown in Fig. 6.  In blue bronze $\Delta\rho/\rho_{ave}$ is enhanced by low-level weak-pinning doping, and reduced by heavily weak-pinning doping or by a strong-pinning impurity \cite{Xue_1996}.  Perhaps the non-linear behavior of $\Delta\rho/\rho_{ave}$ with $RRR$ reflects a crossover from one pinning regime to another. 

\medskip Last, we observe that single crystal sample 3 shows a feature at $T_1$, but nothing obvious associated with either $T_2$ or $T_3$ ({\it e.g.} thermal hysteresis is not visible in Fig. 7).  Sample 3 is the very thin crystal we used to measure the Hall effect.  If the sharpness of the transitions correlates with $RRR$, then this crystal should show features intermediate between those of samples 1 and 2.  We might speculate that the crystal was damaged by the grinding process but, if so, why is its $RRR$ so high relative to most of our other samples and why does subsequent bending degrade it so?  The presence (or absence) of density wave domain walls and their interactions with the surface of the sample may play a role here.  This is clearly a subject requiring further study.

\subsection{Magnetoresistivity}

The only previous measurement of the magnetoresistivity of \a-U that we are aware of focused on the anisotropy at 4.2 K \cite{Berlincourt_1959}.  In this earlier study, the magnetoresistivity $\Delta\rho/\rho(0)$ varied from a little more than 10\% to a little less than 30\% at 3 T and 4.2 K (depending upon the relative orientation of current and field).  By comparison, the magnetoresistivity of sample 2 is just over 90\% at 3 T and 4.2 K, consistent with a higher quality sample.  The failure of Kohler scaling is not surprising since the CDW states modify the Fermi surface.  What does strike us as surprising, however, is that the low temperature data Kohler-scales rather well with the high temperature data for both orientations studied (see Fig. 9).  It is also interesting that Kohler scaling fails in a different way in polycrystal sample 5 (see Fig. 10).  This suggests that there may be a strong anisotropy in the magnetoresistance, much stronger than the roughly 20\% anisotropy we observe for fields parallel and perpendicular to the $c$-axis at 2 K in high fields.

\subsection{Hall Effect}

\noindent Previous measurements of the Hall effect by \cite{Berlincourt_1959} (polycrystal samples with $RRR$'s of 7 and 12) and by \cite{Cornelius_1980} (a single crystal with the magnetic field parallel to the $b$-axis and $RRR$ = 11) show a behavior qualitatively similar to our results.  At 5 K we find that the magnitude of the Hall coefficient near 2 T is about 13 times larger than that reported by \cite{Cornelius_1980} while the Hall coefficient reported by \cite{Berlincourt_1959} has barely changed sign.  Perhaps a more fundamental difference is that we find a low temperature Hall coefficient whose magnitude increases up to about 6 T, after which its magnitude decreases with increasing field strength.  Although \cite{Cornelius_1980} only show data at 0.6 T and 1.8 T, the trend is clearly towards a smaller magnitude, opposite to our results. While these differences may be due to sample quality they may reflect an anisotropy of the Hall coefficient. 

\medskip Last, we remind the reader that the data comprising Figs. 11 and 12 were acquired both before and after the bending of the sample described above.  Although the resistivity was clearly affected (see Fig. 7), the bending did not change the Hall coefficient within experimental uncertainty.  This is a clear indication that the unusual temperature and field dependence of the Hall coefficient reflects fundamental changes in Fermi surface topology due to the CDW transitions and not changes in the electronic relaxation time(s).

\subsection{The Spin Density Wave Hypothesis}

Several of our experimental observations are consistent with a spin density wave (SDW) state coexising with the CDW.  Such a coexistence is not unusual, for example the two states coexist in chromium where a CDW is ``created'' by the dominant SDW state through electron-phonon coupling \cite{Fawcett_1988}.  Evidence for the existence of an SDW in \a-U includes:  A field dependence to the shape of the resistive transitions, especially at $T_2$ (see Fig. 2) and at $T_3$ (see Fig. 3), and a field dependent magnetic hysteresis below $T_3$ (see Fig. 5).   The negative curvature of the magnetoresistance observed for temperatures near $T_3$ may reflect the high field suppression of magnetic scattering from spin fluctuations associated with an SDW.  Lastly, the ``freezing out'' of magnetic scattering associated with an SDW may contribute to the general resistive drop observed as the CDW phase transitions are crossed (as opposed to the increase one would expect from the gapping of the Fermi surface).  On the other hand, one might expect a more dramatic anisotropy in the magnetoresistance than that shown in Fig. 8 if magnetic anisotropy and/or SDW's were present. 

\subsection{Spherical Fermi Surface Approximations}

\noindent If we assume a free-electron-like spherical Fermi surface for low temperature \a-U we can estimate some fundamental quantities as follows:  Since the ``zero field'' Hall coefficient appears to be saturating at the lowest temperatures we extrapolate its value at 5 K to $H = 0$ (see the inset of Fig. 12) and use $r_H = -1/ne$ to estimate $n = 1.1{\times}10^{28}$ carriers/m$^3$ (about 0.2 electrons/U-atom). We use this result to calculate a free electron specific heat coefficient $\gamma_{free}$ = 2.0 mJ/mol-K$^2$ which, combined with the measured value of $\gamma$ reported by \cite{Lashley_2001}, yields a thermal effective mass m$^*$ = 4.5 m$_e$ (where m$_e$ is the bare electron mass).  This value of m$^*$ is relatively large for an element \cite{Kittel}.  We then combine these quantities in the well known Drude relationship

\begin{equation}
\rho = \frac{m^*}{ne^2\tau}, \label{eq:Drude}
\end{equation}

\medskip\noindent where $e$ is the electronic charge and $\tau$ is the relaxation time, then we use $l=v_F\tau$ where $v_F$ is the Fermi velocity and $l$ is the mean free path to get

\begin{equation}
\rho{l} = \left(\frac{3\pi^2}{n^2}\right)^{1/3} \frac{\hbar}{e^2}. \label{eq:mfp}
\end{equation}

\medskip\noindent Single crystal sample 1 has a residual resistivity $\rho_o$ = 0.33 $\mu\Omega$-cm, so Eqn. \ref{eq:mfp} yields $l_1$ = 8000 $\stackrel{\rm o}{\rm A}$.  Similarly, polycrystal sample 5 has $\rho_o$ = 2.6 $\mu\Omega$-cm, the largest residual resistivity of the samples we have studied, giving $l_5$ = 1000 $\stackrel{\rm o}{\rm A}$.  

\subsection{Superconductivity}

\noindent  Despite critical temperatures that vary by just over a factor of 2 and upper critical fields that vary by just under a factor of 3, the shapes of the two critical field curves are identical within experimental uncertainty (see Fig. 14) suggesting that the superconducting states are also identical.  Since single crystal sample 4 was grown at temperatures below the $\alpha-\beta$ phase transition it cannot contain impurity phases of either $\beta$-U or $\gamma$-U.  Ignoring the possibility that the superconductivity observed by so many researchers in so many samples over so many years is due to the same kind of insidious ubiquitous impurity, we take the results embodied in Fig. 14 as unambiguous experimental proof that the superconductivity of \a-U is intrinsic.

\medskip The positive curvature in the upper critical field of \a-U is inconsistent with the conventional theory of BCS, Type II superconductivity \cite{WHH_1966,Maki_1966}.  Indeed, positive curvature in \Hc2(T) has become a ``common'' attribute of unconventional superconductivity, having been observed in heavy fermion superconductors \cite{DeLong_1987}, organic superconductors \cite{Lee_1997}, borocarbide superconductors \cite{gms_2001}, and cuprate superconductors \cite{Ando_1990} for example.  For our samples, the positive curvature causes H$_{c2}$(0) to dramatically exceed the orbital limit given by

\begin{equation} 
H_{c2}^* = 0.693(-(dH_{c2}/dT)|_{T_c})T_c, 
\end{equation}

\medskip\noindent \cite{Hake_1967}; though it does not exceed the paramagnetic, or Chandrasekhar--Clogston limit given by

\begin{equation}
H_{po} = 1.84 T_c ({\rm tesla}),
\end{equation}

\medskip\noindent \cite{Chandrasekhar_1962,Clogston_1962}.  The numerical values of H$_{c2}^*$ and H$_{po}$ for our samples are given in Table 2.

\medskip As we proceed to analyze our \Hc2(T) data the reader is cautioned that we are applying conventional Type II theory (except where noted) to an unconventional curve.  The reader is also cautioned that critical fields determined from transport measurements can differ from those resulting from thermodynamic measurements (that is, the positive curvature may be an artifact).  The upper critical field is related to the coherence length $\xi$ by the well known relation

\begin{equation}
H_{c2} = \frac{\Phi_o}{2\pi\xi^2},
\end{equation}

\medskip\noindent where $\Phi_o$ is the flux quantum.  Calculated values of $\xi$ may be found in Table 2.  For both samples $\xi$ is much less than our estimates of the transport mean free path $l$, placing these \a-U samples in the clean limit. 

\medskip \cite{Toxen_1965} has noted a phenomenological relationship, which works very well for several other elemental superconductors (including the Type II superconductor Nb), relating the scaled slope of the upper critical field near \Tc to the superconducting coupling parameter

\begin{equation}
\frac{T_c}{H_{c2}(0)}\left|\frac{dH_{c2}(T)}{dT}\right|_{T_c} = \frac{\Delta(0)}{k_BT_c},
\end{equation}

\medskip\noindent where $\Delta(0)$ is the magnitude of the superconducting gap at $T=0$.  The values of this quantity, about 1.1 (quite a bit less than the BCS value of 1.73), are given in Table 2.    

\medskip Many of the experimental results discussed above are consistent with a two-band model \cite{Suhl_1958,Moskalenko_1959,Geilikman_1974} for the superconductivity of \a-U.  In this model the Fermi surface passes through two (or more) distinct bands each characterized by its own density of states, anisotropy, etc.  Each band has its own set of Cooper pairs and energy gaps. The bands are coupled through phonon emission and absorption processes.  (The existence of at least two bands near the Fermi surface of \a-U is clearly demonstrated by the Hall effect measurements discussed above.)  Within this model one can show that one gap must be larger than the BCS prediction and the other must be smaller.  It is the smaller gap that dominates the thermodynamic properties at low temperature (such as H$_{c2}$).  Extensions of this model have described the positive curvature in \Hc2(T) for some of the borocarbide superconductors  for example \cite{Shulga_1998}.  A detailed comparison of our upper critical field data with the predictions of this model will appear elsewhere.

\subsection{General}

\medskip We remind the reader that the measurements presented above were all made on samples cut from the original crystalline shards.  It is, of course, very important that this new approach to sample fabrication be fully developed to produce a new generation of high quality \a-U single crystals. The issue of anisotropy in structural, thermodynamic, and transport properties such as those reported here must be fully explored.  A more quantitative analysis of these quantities must await a clearer picture of the Fermi surface of \a-U than presently exists.  

\section{Conclusions}

We have presented and discussed the electrical resistivity, magnetoresistance, and Hall effect in four single crystal samples and one polycrystalline sample of \a-U.  The $RRR$'s of these samples vary from 13 to 315, about 10 times larger than for any sample in the literature of which we are aware, indicating very high quality samples.  Matthiessen's law appears to hold above the onset of the CDW transitions which begin at 43 K, but it does not appear to hold below this temperature.  Sharp features at all three charge CDW transitions were observed and the effects of high magnetic fields were characterized.  The magnetoresistance is anisotropic, reaches 1000\% at 2 K and 18 T, and does not exhibit Kohler scaling.  The Hall coefficient is positive, independent of magnetic field, and slightly temperature dependent above about 40 K in agreement with earlier studies.  Below 40 K the Hall coefficient varies with field, changes sign, and becomes much more strongly negative than has been seen previously.  Some of these results suggest that a spin density wave coexists with the charge density wave over a portion of the phase diagram.  Superconductivity has been observed in two of our samples and we argue that it is intrinsic to \a-U.  The upper critical field exhibits positive curvature which suggests that the superconductivity is unconventional, and we discussed the possible application of the two-band model to this material.  We also estimated several parameters characterizing the transport and superconductivity of \a-U.

\section{Acknowledgments}

\noindent We gratefully acknowledge the comments and suggestions of V.~Z. Kresin and L.~E. De Long.  This work was supported by the National Science Foundation under DMR--0071947 and DMR--0305397 (GMS).  Some of the measurements were made at the National High Magnetic Field Laboratory which is supported by the National Science Foundation and the State of Florida.  This work was partially supported by the Director for Energy Research, Office of Basic Energy Sciences of the U.~S. Department of Energy.

\vfill\newpage
      
\noindent{\bf FIGURE CAPTIONS}

\bigskip

\bigskip\noindent Figure 1.  The reduced resistivity ($\rho(T)/\rho($300 K) of two single crystals and one polycrystal sample of \a-U in zero field.  The samples span our experimental range of residual resistivity ratios ($RRR$).  Figure 1a shows the data from 2 K to 300 K. Figure 2b shows the data below 50 K, the data for samples 1 and 5 have been vertically shifted by the amounts shown in parentheses for clarity.

\bigskip\noindent Figure 2. The resistivity of \a-U below 50 K in magnetic fields $H =$ 0, 10 T, and 18 T, applied along the $c$--axis as shown.  The charge density wave phase transitions occur at $T_1$, $T_2$ and $T_3$.  

\bigskip\noindent Figure 3.  The first derivative of the temperature dependent resistance of single crystal sample 2 in zero field (solid line) and in a field of 18 T applied along the $c$--axis (dotted line).  The arrows denote the direction of temperature change.

\bigskip\noindent Figure 4.  The low temperature, zero field resistivity of polycrystal sample 5 is shown in Figure 4a where the direction of changing temperature is denoted with arrows.  Figure 4b shows the derivative of the resistivity.  Data taken with the temperature increasing (decreasing) is represented by the solid (dotted) line.

\bigskip\noindent Figure 5. The relative magnitude of the thermal hysteresis in the resistivity of single crystal sample 1 in magnetic fields of zero (solid line), 10 T (dashed line), and 18 T (dotted line).  The field is applied along the 
$c$--axis.

\bigskip\noindent Figure 6.  The relative magnitude of the thermal hysteresis in the zero field resistivity of single crystal sample 2 (solid line), single crystal sample 1 (dotted line), and polycrystal sample 5 (dashed line).

\bigskip\noindent Figure 7.  The zero field resistivity of single crystal sample 3 (a) and its derivative (b) are shown before (solid line) and after (dotted line) bending (see text).

\bigskip\noindent Figure 8. The magnetoresistivity of single crystal sample 2 with fields applied (a) parallel and (b) perpendicular to the $c$--axis at the temperature shown.  Most of the data sets are scaled by the factors shown in parentheses.

\bigskip\noindent Figure 9.  Kohler scaled (see text) magnetoresistivity of single crystal sample 2 with fields applied (a) parallel and (b) perpendicular to the $c$--axis at the temperature shown.

\bigskip\noindent Figure 10.  The magnetoresistivity (a) and Kohler scaled magnetoresistivity (b) of polycrystal sample 5 at the temperatures shown.

\bigskip\noindent Figure 11.  The Hall resistance of single crystal sample 3, with magnetic fields applied along the 
$c$--axis at the temperatures shown.  Figure 11a shows representative data over the full range of temperatures and fields (note that the 5 K data has been reduced by a factor of 3).  Figure 11b shows the Hall resistance at low fields in the vicinity of 22 K where the Hall coefficient changes sign near zero field.

\bigskip\noindent Figure 12.  The Hall coefficient of single crystal sample 3 plotted as a function of temperature at the fields shown (applied parallel to the $c$--axis).  The higher temperature data (solid squares) were derived from the slopes of the isothermal Hall resistance measured as a function of magnetic field.  The solid lines are guides to the eye.  The inset shows the field dependence of the Hall coefficient at 5 K, the solid line is a fit to the data (see text).

\bigskip\noindent Figure 13.  The zero-field temperature dependence of the resistance of single crystal sample 4 and polycrystal sample 5 at low temperature.  Phase transitions are marked with arrows (see text).  The inset shows the isothermal resistance of polycrystal sample 5 as a function of magnetic field; phase transitions are again marked with arrows (see text).

\bigskip\noindent Figure 14. The upper critical magnetic field of single crystal sample 5 (closed circles, with fields applied along the $c$-axis) and polycrystal sample 5 (open circles).  The data are plotted as $T/T_c$ vs. $H_{c2}(T)/H_{c2}(0)$.  The solid line is a guide to the eye.

\bigskip\noindent Figure 15.  A phase diagram in the $H$-$T$ plane showing the position of the ``3 K feature'' appearing in polycrystal sample 5.  The solid line is a fit to the data (see text).  The inset shows the temperature dependent resistance of this sample at 0.5 T (large enough to suppress superconductivity), the position of the feature is identified with an arrow.

\vfill\newpage

\bigskip\noindent\centerline{\bf Table 1.}

\bigskip\noindent\centerline{\begin{tabular}{|c|c|c|c|c|c|c|} \hline
sample \# & $RRR^d$ & $n^e$ & $n^f$ & T$_{nFL}^g (K)$ & T$_{nFL}^h$ (mK) & Reference \\ \hline\hline
1     & 117 & 0.71 & 3.1 & 3.2 & 77 & $i,j$ \\ \hline
2     & 315 & 0.62 & 3.5 & 55 & 100 &  \\ \hline
3$^a$ & 183 & 0.74 & 3.3 & 40 & 67 &  \\ \hline
3$^b$ & 29  & 0.85 & 3.1 & .34 & 45 &  \\ \hline
4     & 66   &   & &  &  & $j$ \\ \hline   
5$^c$ & 13  & 0.73 & 3.0 & .13 & 32 &  \\ \hline
\end{tabular}}

%\Tc's are 90% onsets to be consistent with Lashley PRB.

\bigskip Some parameters (defined in the text) characterizing the zero field resistivity of the \a-U samples described in this paper.  All samples are single crystals unless noted.

\bigskip$^a$ Sample used to measure Hall effect.

\medskip$^b$ Hall sample after plastic deformation.

\medskip$^c$ Polycrystal.

\medskip $^d$ Residual resistivity ratio: $R$(T$\sim$300 K) / $R$(T$\sim$0).

\medskip $^e$ Power from fit to $\rho(T)$ = $\rho_o$ + $AT^n$ for 60 K $\leq T \leq$ 200 K as described in text.

\medskip $^f$ Power from fit to $\rho(T)$ = $\rho_o$ + $AT^n$ for 4 K $\leq T \leq$ 10 K as described in text.

\medskip $^g$ T$_{nFL}$ derived from fit to $\rho(T)$ = $\rho_o$ + $AT^n$ for 60 K $\leq T \leq$ 200 K as described in text.

\medskip $^h$ T$_{nFL}$ derived from fit to $\rho(T)$ = $\rho_o$ + $AT^3$ for T $\leq$ 7 K as described in text.

\medskip $^i$ \cite{JLS1}

\medskip $^j$ \cite{Lashley_2001}

\vfill\newpage

\bigskip\noindent\centerline{\bf Table 2.}

\bigskip\noindent\centerline{\begin{tabular}{|c|c|c|} \hline
Quantity           & Sample 4   & Sample 5       \\ \hline\hline
$T_c$(K)           & .819 ${\pm}$ .013  & 1.761 ${\pm}$ .005    \\ \hline
$H_{c2}$(0) (mT)   & 81.6 ${\pm}$ 1    & 228 ${\pm}$ 7         \\ \hline
$\xi$ ($\stackrel{\rm o}{\rm A}$) & 
 635 $\pm$ 4       & 380 $\pm$ 6           \\ \hline
$-(dH_{c2}/dT)|_{T_c}$ (mT/K) & 98.2 ${\pm}$ 10   & 148 ${\pm}$ 3         \\ \hline
$ T_c/H_{c2}(0)\enspace (dH_{c2}/dT)|_{T_c}$
                   & 0.99 ${\pm}$ .13  & 1.14 ${\pm}$ .06      \\ \hline
H$_{c2}^*$ (mT)    &  56 ${\pm}$ 3  & 180 ${\pm}$ 4 \\ \hline
H$_{po}$  (T)      &  1.5  & 3.2  \\ \hline
\end{tabular}}

%\Tc's are 90% onsets to be consistent with Lashley PRB.

\bigskip Superconducting properties of \a-U single crystal Sample 4 and polycrystal Sample 5.  The quantities listed are defined in the text.

\vfill\end{document}